\documentclass[prb,superscriptaddress,floats,floatfix,]{revtex4}

\usepackage{graphicx}
\usepackage{graphics}
\usepackage{gensymb}
\usepackage{amsmath}
\usepackage{bm}
\usepackage{newfloat}
\DeclareFloatingEnvironment[name={Supplementary Figure}]{suppfigure}

\begin{document}

\title{Spin-valley lifetimes in a silicon quantum dot with tunable valley splitting}

\author{C. H. Yang}
\author{A. Rossi}
\email{a.rossi@unsw.edu.au}
\affiliation{Australian Research Council Centre of Excellence for Quantum Computation and Communication Technology, School of Electrical Engineering \& Telecommunications, The University of New South Wales, Sydney 2052, Australia}
\author{R. Ruskov}
\affiliation{Laboratory for Physical Sciences, 8050 Greenmead Dr., College Park, MD 20740, USA}
\author{N. S. Lai}
\author{F. A. Mohiyaddin}
\affiliation{Australian Research Council Centre of Excellence for Quantum Computation and Communication Technology, School of Electrical Engineering \& Telecommunications, The University of New South Wales, Sydney 2052, Australia}
\author{S. Lee}
\affiliation{Network for Computational Nanotechnology, Birck Nanotechnology Center, Purdue University, West Lafayette, IN 47907, USA}
\author{C. Tahan}
\affiliation{Laboratory for Physical Sciences, 8050 Greenmead Dr., College Park, MD 20740, USA}

\author{G. Klimeck}
\affiliation{Network for Computational Nanotechnology, Birck Nanotechnology Center, Purdue University, West Lafayette, IN 47907, USA}
\author{A. Morello}
\author{A. S. Dzurak}
\affiliation{Australian Research Council Centre of Excellence for Quantum Computation and Communication Technology, School of Electrical Engineering \& Telecommunications, The University of New South Wales, Sydney 2052, Australia}

\date{\today}

\begin{abstract}
Although silicon is a promising material for quantum computation, the degeneracy of the conduction band minima (valleys) must be lifted with a splitting sufficient to ensure formation of well-defined and long-lived spin qubits. Here we demonstrate that valley separation can be accurately tuned via electrostatic gate control in a metal-oxide-semiconductor quantum dot, providing splittings spanning 0.3 - 0.8 meV. The splitting varies linearly with applied electric field, with a ratio in agreement with atomistic tight-binding predictions. We demonstrate single-shot spin readout and measure the spin relaxation for different valley configurations and dot occupancies, finding one-electron lifetimes exceeding 2 seconds. Spin relaxation occurs via phonon emission due to spin-orbit coupling between the valley states, a process not previously anticipated for silicon quantum dots. An analytical theory describes the magnetic field dependence of the relaxation rate, including the presence of a dramatic rate enhancement (or hot-spot) when Zeeman and valley splittings coincide. 
\end{abstract}

\maketitle

\section{Introduction}
\label{intro}
Silicon is at the heart of all modern microelectronics. Its properties have allowed the semiconductor industry to follow Moore's law for nearly half a century, delivering nowadays billions of nanometre-scale transistors per chip. Remarkably, silicon is also an ideal material to manipulate quantum information encoded in individual electron spins~\cite{kane,maune,pla}. This is a consequence of the weak spin-orbit coupling and the existence of an abundant spin-zero isotope, which can be further enriched to obtain a ``semiconductor vacuum'' in which an electron spin can preserve a coherent quantum superposition state for exceptionally long times~\cite{tyry}.\\\indent
In order to define a robust spin-$^1$/$_2$ qubit Hilbert space, it is necessary that the energy scale of the two-level system is well separated from higher excitations. In this respect, a major challenge for the use of silicon is represented by the multi-valley nature of its conduction band. In a bulk silicon crystal the conduction band minima are six-fold degenerate, but in a two-dimensional electron gas (2DEG), the degeneracy is broken~\cite{Ando1982} into a two-fold degenerate ground state ($\Gamma$ valleys) and a four-fold degenerate excited state ($\Delta$ valleys), due to vertical confinement of electrons with different effective mass along the longitudinal and transverse directions, respectively. Furthermore, the $\Gamma$ valley degeneracy is generally lifted by a sharp perpendicular potential~\cite{sham,boykin,Saraiva2009, Saraiva2011} and the relevant energy separation is termed the valley splitting (VS).\\\indent
The valley splitting depends on physics at the atomic scale~\cite{andosolo,chutia,fec} (e.g. roughness, alloy and interface disorder), and so it is not surprising that experiments have revealed a large variability of splittings among devices, ranging from hundreds of $\mu$eV~\cite{Ando1982,goswami,lansbergen,roche} up to tens of meV in exceptional cases~\cite{taka}. At present, the lack of a reliable experimental strategy to achieve control over the VS is driving an intense research effort for the development of devices that can assure robust electron spin qubits by minimising multi-valley detrimental effects~\cite{Culcer2009,li}, or even exploit the valley degree of freedom~\cite{lithium,Culcer2012} for new types of qubits.\\\indent
Another crucial parameter to assess the suitability of a physical system to encode spin-based qubits is the relaxation time of spin excited states ($T_1$). Spin lifetimes have been measured for gate-defined Si quantum dots (QDs)~\cite{Xiao2010}, Si/SiGe QDs~\cite{Simmons2011,hayes} and donors in Si~\cite{Morello}, reporting values which span from a few milliseconds to a few seconds. Furthermore, the dependence of the spin relaxation rate ($T_1^{-1}$) on an externally-applied magnetic field ($B$) has been investigated. Different mechanisms apply to donors and QDs, accounting for observed $T_1^{-1}\propto B^5$ and $B^7$ dependencies~\cite{floris}, respectively. In principle, $T_1^{-1}(B)$ depends on the valley configuration and the details of the excited states above the spin ground state. However, until now, no experimental observation of the effects of a variable VS on the relaxation rate has been reported.\\\indent
Here, we demonstrate for the first time that the valley splitting in a silicon QD can be finely tuned by direct control of an electrostatic gate potential. We find that the dependence of the VS on vertical electric field at the Si/SiO$_2$ interface is strikingly linear, and show that its tunability is in excellent agreement with atomistic tight-binding predictions. We demonstrate accurate control of the VS over a range of about 500 $\mu$eV and use it to explore the physics of spin relaxation for different QD occupancies ($N$=1, 2, 3). We probe both the regime where the VS is much larger than the Zeeman splitting at all magnetic fields and that where the valley and spin splittings are comparable. We observe a dramatic enhancement of the spin decay rate (relaxation hot-spot) when spin and valley splittings coincide. To our knowledge, such hot-spots have been predicted for relaxation involving orbital states~\cite{firstHS,raith} (not valley states), but these are yet to be observed. We develop an analytic theory that explains the $B$-field dependence of the relaxation rates and the details of the relaxation hot-spot in terms of admixing of spin-valley states. This mechanism is seen to be significantly more prominent than the conventional spin-orbit hybridization~\cite{hanson}.
\begin{figure}[]
\includegraphics[scale=0.45]{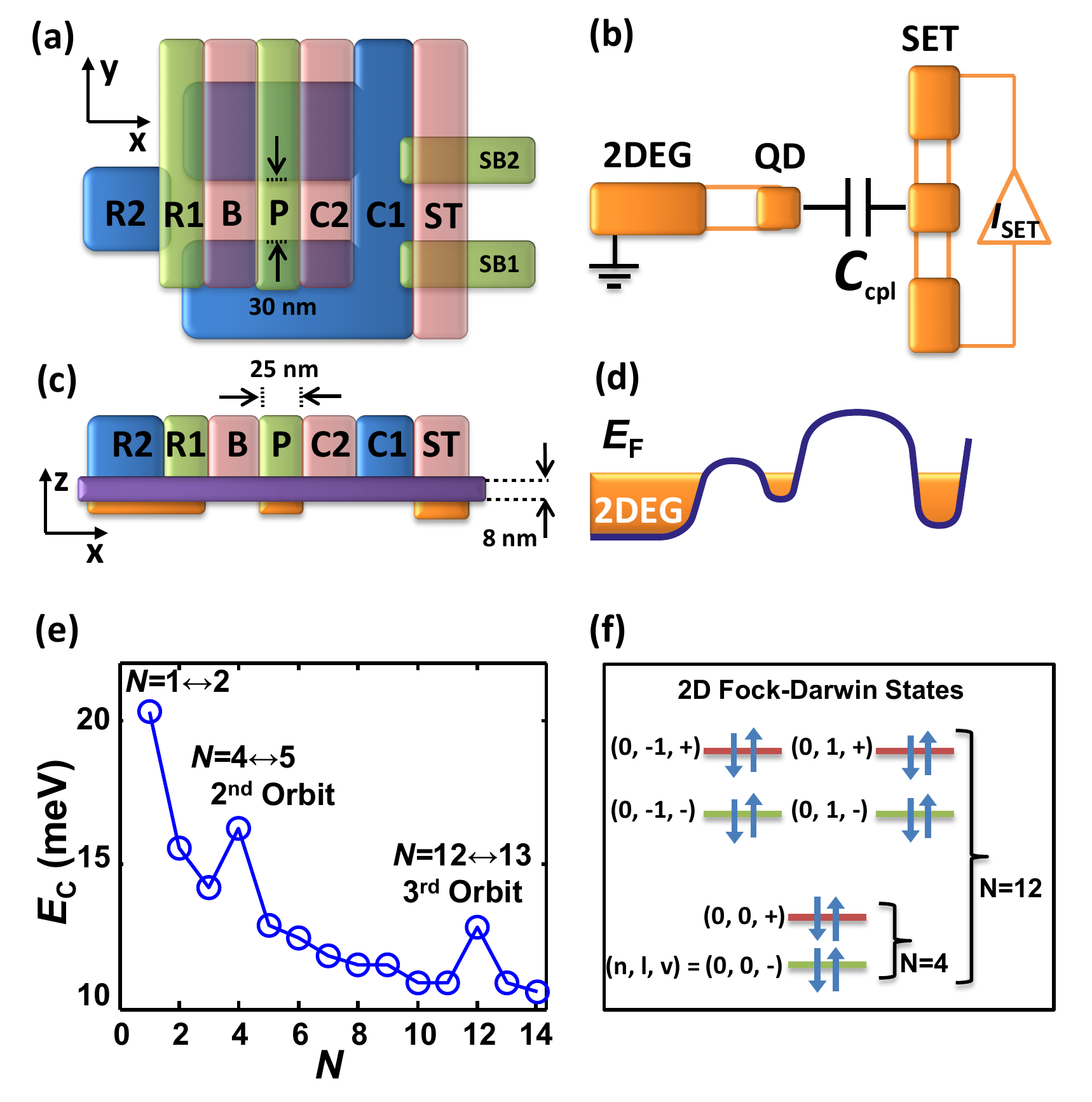}
\caption{Device architecture and addition energy spectrum. (a) Schematic (top view) of the device's gate layout. Different colours represent different layers within the gate stack. (b) Schematic diagram of the single-lead QD (left) and SET detector (right). Regions where an electron layer is formed are coloured in orange. The readout signal ($I_\textup{SET}$) is sensitive to the QD charge state due to the QD/SET capacitive coupling ($C_\textup{cpl}$). (c) Device cross-sectional schematic. An electron layer is formed underneath the positively biased gates: R1 and R2 define the QD reservoir; P controls the QD population; and ST the sensor's island. The SiO$_2$ layer (in purple) thickness and plunger gate width are indicated. (d) Energy diagram showing qualitatively the conduction band profile in the device. Electrons accumulate wherever the gate bias lowers the conduction band below the Fermi level, $E_\textup{F}$. (e) Charging energy as a function of electron number. Spikes corresponding to complete 2D shell filling are observed. (f) Schematic of electron filling for two-valley 2D Fock-Darwin states. Each state can hold two electrons of antiparallel spin and is identified by a pair of quantum numbers ($n,l$) and its valley occupancy ($v$).}
\label{device}
\end{figure}
\section{Results}
\subsection{QD Addition Spectrum}
\label{spectrum}
Our device is fabricated using a multi-level gated metal-oxide-semiconductor (MOS) technology~\cite{Angus2007}, and its architecture is depicted in Fig.~\ref{device}(a)-(d). A quantum dot is formed under gate P by applying a positive bias voltage to induce an electron accumulation layer. Strong planar confinement for the dot's potential well is achieved by negatively biasing gates B, C1 and C2. A 2DEG reservoir is also induced by positively biasing gates R1 and R2, and the QD occupancy can be modified by inducing electrons to tunnel between this reservoir and the dot. The remaining gates, namely SB1, SB2, ST, are employed to define a single-electron transistor (SET), capacitively coupled to the QD and used as a read-out device. The high flexibility of our design would allow us to use the same device also as a (single-lead) double dot structure by rearranging the gate bias (e.g. dots can be formed under gates B and C2). However, in this work, we only present results relevant to the single-dot configuration.\\\indent
In order to characterize the addition spectrum of the QD, we make use of a technique previously developed for GaAs-based systems which combines charge detection and gate pulsing~\cite{Elzerman2004}. There is no direct transport through the single-lead QD and, therefore, addition/removal of charge  is only detected via modifications in the SET current. In particular, charge transitions are detected as current peaks in the SET signal whenever the QD energy eigenstates come into resonance with the reservoir's Fermi level. Note that the SET-QD coupling is merely capacitive (via $C_\textup{cpl}$) and electrons do not tunnel between them. In order to maximize charge sensitivity in the detector, we employ dynamic voltage compensation~\cite{Yang2011} on different gates which makes our read-out signal virtually unaffected by slow charge drifts and random charge rearrangements.  A comprehensive discussion of the charge stability measurements can be found in the Supplementary Note 1.\\\indent 
Fig. 1(e) illustrates the addition energy spectrum for the first fourteen electron additions to the QD. There is very little variation of charging energy ($E_\textup{C}$) for high occupancies ($E_\textup{C}\approx$~11 meV for $N>9$). However, by decreasing the electron number, the charging energy steadily increases, as expected when the dot size is significantly affected by the electron number. This evidently indicates that the few-electron regime has been achieved. Most interestingly, the energy spectrum shows peaks for the addition of the 5th and 13th electrons. The extra addition energy needed for those transitions can be attributed to complete filling of the first and second orbital shells. As illustrated in Fig. 1(f), this is consistent with the energy spectrum of two-valley 2D Fock-Darwin states~\cite{kou}, where the first and second orbital shells hold 4 and 8 electrons respectively. This confirms that we can probe the occupancy until the last electron. To our knowledge, such a clear manifestation of two-dimensional shell structure has been observed before only in InGaAs dots~\cite{tarucha} and in Si/SiGe dots~\cite{Borselli2011}.
\begin{figure*}[]
\includegraphics[scale=0.5]{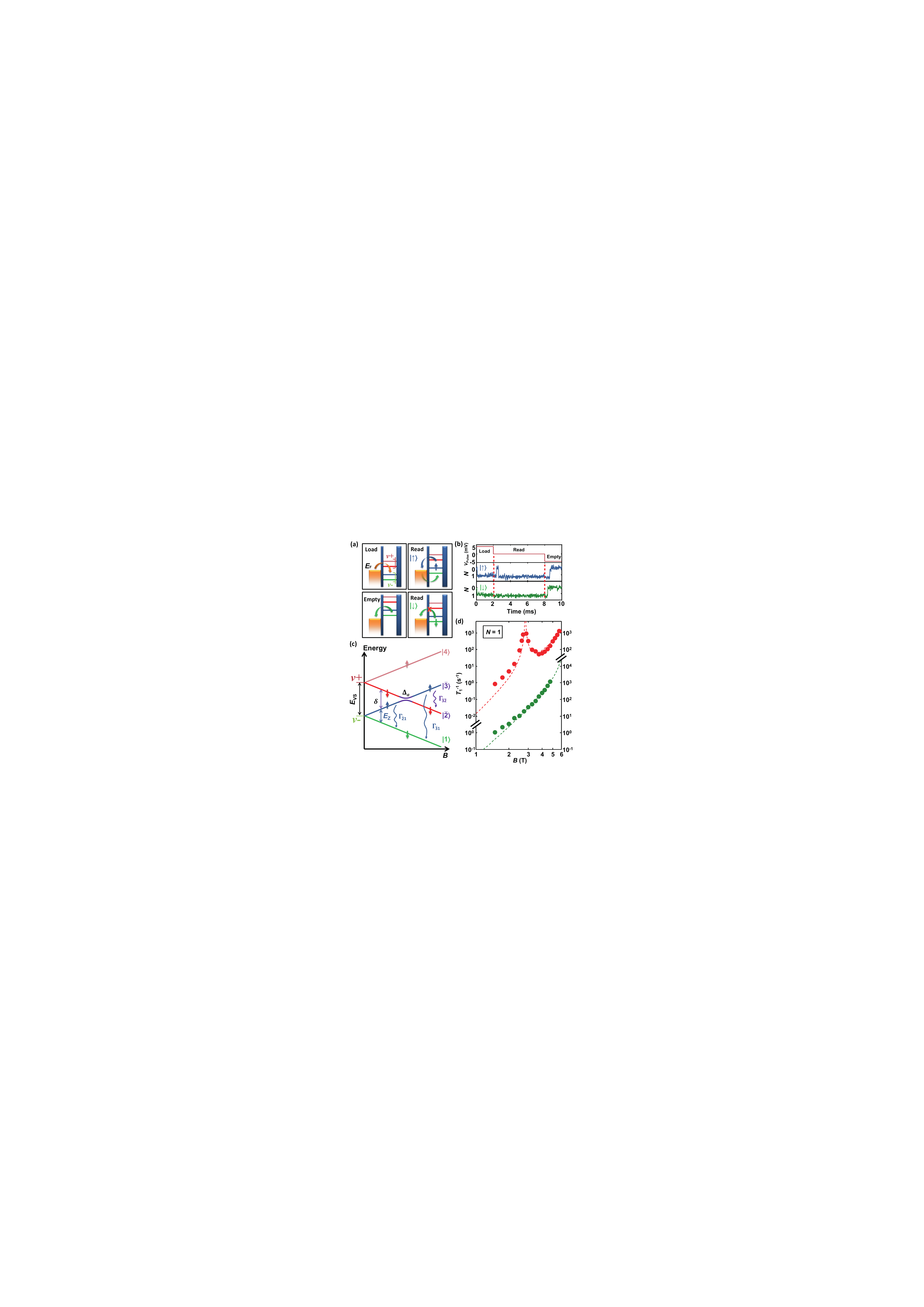}
\caption{Spin readout and relaxation rates for single-electron occupancy. (a) Schematic diagram showing the effect of the 3-level pulse sequence on the electro-chemical potential of the dot. Energy levels in the QD are Zeeman split according to spin polarisation and valley degeneracy is lifted. For clarity, only lower valley states are shown to be loaded/unloaded. (b) Pulsing sequence (top) for the single-shot spin readout and normalised SET signal for spin-up (middle) and spin-down (bottom). (c) Energy diagram of the 1-electron spin-valley states as a function of $B$-field. Maximum mixing of spin and valley degrees of freedom occurs at the anticrossing point where Zeeman and valley splittings coincide. Relevant relaxation processes are sketched. (d) Relaxation rates as a function of magnetic field for different valley splittings. Data points for $E_\textup{VS}=0.75$meV, $E_\textup{VS}=0.33$meV are shown as green and red circles, respectively. Dashed lines are the calculated relaxation rates fitted with $r$=1.7~nm (green), $r$=1.1~nm (red).}
\label{T1vsB}
\end{figure*}
\subsection{Spin-Valley Lifetimes}
\label{lifetime}

In order to measure the spin state of individual electrons in the QD, we use an energy-selective readout technique~\cite{Elzerman2004b}. The readout protocol consists of three phases clocked by a three-level pulsed voltage applied to gate P, which directly controls the dot's electrochemical potential [see Fig.~\ref{T1vsB} (a)]. Firstly, an electron of unknown spin is loaded into the dot causing a sudden decrease in the sensor current. Next, the potential of the dot is lifted so that the Fermi level of the reservoir lies between the spin-up and spin-down states of the dot, meaning that a spin-up electron can tunnel off the dot while a spin-down electron is blocked. This is the read phase, during which the presence of a spin-up state would be signalled by a current transient (spin-up tunnels out and then spin-down tunnels in) whereas a spin-down electron would lead to no current modification. Finally, the dot's potential is further lifted to allow the electron to tunnel off, regardless of its spin orientation. In Fig.~\ref{T1vsB} (b) single-shot traces for both spin-up (in blue) and spin-down (in green) detection are plotted. The longer the system is held in the load phase before performing a read operation, the more likely it is for the spin-up excited state to decay to the spin-down ground state. Thus, by varying the length of the load phase and monitoring the probability of detecting a spin-up electron, we can determine~\cite{Elzerman2004b,Morello} the spin lifetime, $T_1$.  In our experiments the $B$-field is directed along the [110] crystallographic axis. A comprehensive discussion of both the spin-up fraction measurements and the fitting procedure to evaluate $T_1$ is included in the Supplementary Note 2. As shown in Fig.~\ref{T1vsB} (d), we observe a wide range of spin lifetimes as a function of magnetic field, with lifetimes as long as 2.6s at the lowest fields studied, $B$=1.25T. These are some of the longest lifetimes observed to date in silicon quantum dots~\cite{floris}.\\\indent 
A key focus of our experiment was to electrostatically tune the valley energy separation and measure relaxation rates in different valley configurations and QD electron occupancies. \textit{As we show below, our data definitively indicate that excited valley states play a critical role in the spin relaxation processes.} We develop a theory to explain how changes in the valley splitting affects the spin-valley state mixing and leads to the observed relaxation times.\\\indent
As we detail in Section~\ref{vs ctrl}, we have attained accurate gate control of the valley splitting, allowing us to tune it over a range of hundreds of $\mu$eV. This permits us to conduct experiments in regimes where the valley splitting ($E_\textup{VS}$) is either larger or smaller than the Zeeman spin splitting ($E_\textup{Z}$), depending on the magnitude of the magnetic field [see Fig.~\ref{T1vsB} (c)].\\\indent
Fig.~\ref{T1vsB} (d) presents measurements of spin relaxation rates as a function of magnetic field for two valley splitting values at a fixed dot population of $N=1$. We start by examining a configuration where the valley separation is larger than the spin splitting at all fields (green data set). In other words, we operate in a regime for which
\begin{equation}
\label{BVS}
E_\textup{VS}>E_\textup{Z}=g\mu_\textup{B}B
\end{equation}
where $g$ is the electron gyromagnetic ratio, $\mu_B$ is the Bohr magneton and $B$ is the applied in-plane magnetic field.\\\indent
For $E_\textup{VS}$=0.75 meV [green data in Fig.~\ref{T1vsB} (d)], we observe a monotonic increase of the rate with respect to $B$ that becomes increasingly fast as $E_\textup{Z}$ approaches $E_\textup{VS}$. In our experimental conditions ($B$-field parallel to [110]), the $T_1^{-1}\propto B^5$ dependences for known bulk-like mechanisms in silicon~\cite{glavin,tahan1} should not apply, while predicted~\cite{hayes,raith,floris,tahan2} rates $\propto~B^7$ do not explain the experimental data.\\\indent
By decreasing the valley separation to $E_\textup{VS}=$0.33 meV, we can achieve the condition where the Zeeman splitting matches or exceeds the valley splitting. The red data in Fig.~\ref{T1vsB} (d) illustrate the situation where inequality (1) only holds for $B<$~2.8T. When $E_\textup{Z}=E_\textup{VS}$~(i.e. for $B$=2.8T), a spike in the relaxation rate occurs. Relaxation hot-spots have been previously predicted to occur for spin relaxation involving orbital states in single and coupled QDs~\cite{firstHS,raith,stanoPRL,stanoPRB}.\textit{To our knowledge, this is the first experimental observation of such a phenomenon.}\\\indent
In order to understand the relaxation mechanisms, we have developed a model that takes into account the perturbations in pure spin states due to spin-orbit coupling (SOC), yielding  eigenstates which are admixtures of spin and valley states. The four lowest spin-valley states [see Fig.~\ref{T1vsB}(c)] are defined as $\left|1\right\rangle=\left|v_-,\downarrow\right\rangle$, $\left|2\right\rangle=\left|v_-,\uparrow\right\rangle$, $\left|3\right\rangle=\left|v_+,\downarrow\right\rangle$,
$\left|4\right\rangle=\left|v_+,\uparrow\right\rangle$. These states are considered to be only very weakly affected by higher excitations, such as orbital levels which are at least 8 meV above the ground state in our device~\cite{ourPRB}. In the Supplementary Note 3 we detail how mixing to a 2$p$-like orbital state leads to  a $B^7$ dependence in $T_1^{-1}$ and is, therefore, important mainly for high $B$-fields (above the anticrossing point). At lower fields, the prominent mechanism is the spin-valley admixing, which we now discuss in detail.\\\indent
\begin{figure*}[]
\includegraphics[scale=0.5]{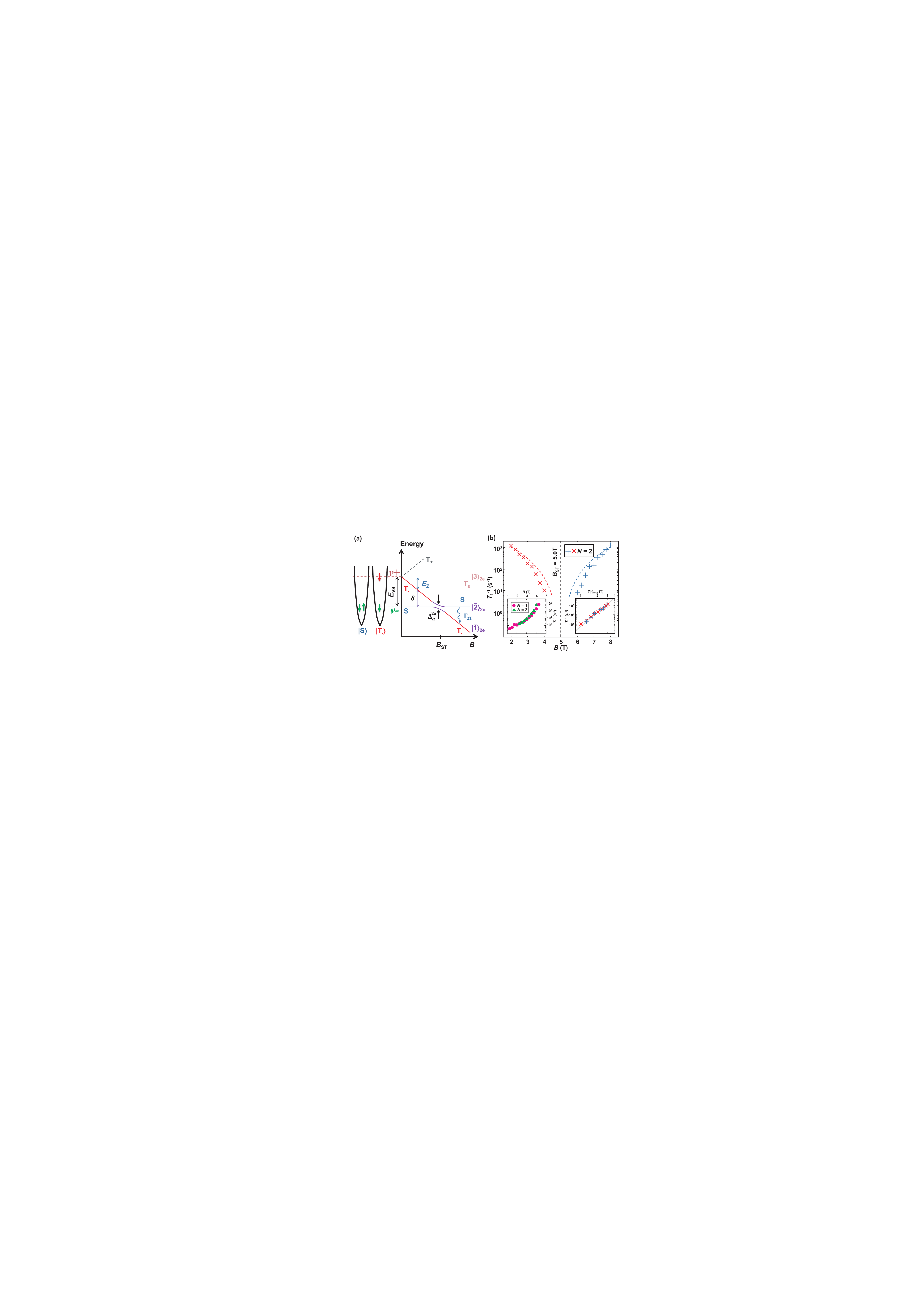}
\caption{Spin-valley relaxation for multi-electron occupancy. (a)  Energy diagrams of the 2-electron spin-valley states in the dot's potential well (left) and as a function of $B$-field (right). Dashed line indicates that $T_+$ is not accessible (see text). (b) Relaxation rate as a function of $B$-field for $N=2$ and $E_\textup{VS}=0.58$meV. Red (blue) crosses represent data points at fields smaller (larger) than the anticrossing point. Dashed lines are the calculated rates fitted with $r^{2e}$=4.76~nm. Right inset: Data from the main graph re-plotted as a function of the modulus of the detuning energy, $\delta$. Points of equal absolute detuning have nearly the same decay rates. $\delta^5$ dependence (grey line) is a guide for the eye. Left inset: Relaxation rate as a function of $B$-field for $N=3$ and $E_\textup{VS}=0.58$meV. Results for $N=1$ are also shown for comparison.}
\label{multi_n}
\end{figure*}
The relaxation between pure spin states is forbidden because the electron-phonon interaction does not involve spin flipping. However, in the presence of interface disorder, SOC can mix states that contain both the valley and spin degrees of freedom, thus permitting phonon-induced relaxation. Indeed, in the non-ideal case of QDs with a disordered interface, roughness can perturb the envelope function of both valleys (otherwise identical for ideal interfaces) and allows one to assume non-zero dipole matrix elements connecting the valley states (see Supplementary Note 3), such as $\bm{r}_{-+}\equiv\langle v_-|\bm{r}|v_+\rangle$, $\bm{r}_{--}\equiv\langle v_-|\bm{r}|v_-\rangle$, $\bm{r}_{++}\equiv\langle v_+|\bm{r}|v_+\rangle$ (for ideal interfaces these are non-zero only due to a strongly suppressed Umklapp process). By means of perturbation theory, we define renormalized excited states $\left|\overline{2}\right\rangle$ and $\left|\overline{3}\right\rangle$ that can relax to the ground state $\left|1\right\rangle$, as they have an admixture of the state $\left|3\right\rangle$ of the same spin projection [see Fig.~\ref{T1vsB}(c)]. The details of the SOC Hamiltonian, $H_\textup{SO}$, and perturbation matrix are reported in the Supplementary Note 3. The leading order wavefunctions are given by:
\begin{equation}
\label{2correct}
\left|\overline{2}\right\rangle^{(0)}=\text{sin}\frac{\gamma}{2}\left|2\right\rangle-\text{cos}\frac{\gamma}{2}\left|3\right\rangle
\end{equation}
\begin{equation}
\label{3correct}
\left|\overline{3}\right\rangle^{(0)}=\text{cos}\frac{\gamma}{2}\left|2\right\rangle+\text{sin}\frac{\gamma}{2}\left|3\right\rangle
\end{equation}
where $\text{cos}\frac{\gamma}{2}\equiv[\frac{1+a}{2}]^{1/2}$, $\text{sin}\frac{\gamma}{2}\equiv[\frac{1-a}{2}]^{1/2}$, and $a\equiv-\delta/\sqrt{\delta^2+\Delta_a^2}$
is an expression involving the detuning from the anticrossing point, $\delta\equiv E_\textup{VS}-E_\textup{Z}$, and the energy splitting at the anticrossing:
\begin{equation}
\Delta_a = 2 |\langle v_-,\!\uparrow\!| H_{SO}|v_+,\!\downarrow\rangle |
= r_{-+} \frac{m_t\, E_{VS}}{\sqrt{2} \hbar}  (\beta_D - \alpha_R)
\label{delta_a-B-(110)}
\end{equation}
where $\beta_D$ ($\alpha_R$) is the Dresselhaus (Rashba) SOC parameter, $\hbar$ is the reduced Planck's constant and $m_\textup{t}=0.198m_\textup{e}$ is the transverse effective electron mass. \\\indent
By evaluating the relaxation rate via the electron-phonon deformation potentials (proportional to the deformation potential constants, $\Xi_{d,u}$), we obtain the rate below the anticrossing as:
\begin{equation}
\label{rate21}
\Gamma_{\bar{2}1}=\text{cos}^2\frac{\gamma}{2}~\Gamma_{v'v}=\frac{\sqrt{\delta^2+\Delta_a^2}-\delta}{2\sqrt{\delta^2+\Delta_a^2}}~\Gamma_{v'v}
\end{equation}
where the pure valley relaxation rates are (for longitudinal and transverse phonons):
\begin{equation}
\Gamma_{v'v}^{(\sigma)}\left[\Delta E_{v'v},\bm{r}\right]  = \frac{\Delta E_{v'v}^5}{4\pi \rho \hbar^6}\, \frac{r^2}{v_{\sigma}^7}\, I^{(\sigma)}
\label{valley-relax-Gam0}
\end{equation}
where $\rho$ is the silicon mass density, $v_\sigma$ is the speed of sound in silicon, $I^{(l)}=4[\frac{\Xi_u^2}{35}+\frac{2\Xi_u\Xi_d}{15}+\frac{\Xi_d^2}{3}]$, 
$I^{(t)}=\frac{16}{105}\Xi_u^2$ are the angular integrals,
and $\Delta E_{v'v}$ and \textbf{\textit{r}} are the energy difference and the dipole matrix element relevant to the transition, respectively (see also Supplementary Note 3). The experimental condition for which the hot-spot occurs (i.e. $E_\textup{VS}=E_\textup{Z}$) is modelled as an anticrossing point for the mixed states $\left|\overline{2}\right\rangle$ and $\left|\overline{3}\right\rangle$. At that point, spin relaxation is maximized and $\Gamma_{\bar{2}1}$ approaches the valley relaxation rate, as $\delta\rightarrow~0$ in eq. (\ref{rate21}).\\\indent
Above the anticrossing (i.e. $E_\textup{VS}<E_\textup{Z}$), the relevant relaxation transitions are $\left|\overline{3}\right\rangle\rightarrow\left|1\right\rangle$ and $\left|\overline{3}\right\rangle\rightarrow\left|\overline{2}\right\rangle$ (the subsequent decay  $\left|\overline{2}\right\rangle\rightarrow\left|1\right\rangle$ is in the form of a fast inter-valley transition and is, therefore, neglected). The analytical formulations of these contributions read:
\begin{equation}
\label{rate31}
\Gamma_{\bar{3}1}=\text{sin}^2\frac{\gamma}{2}~\Gamma_{v'v}=\frac{\sqrt{\delta^2+\Delta_a^2}+\delta}{2\sqrt{\delta^2+\Delta_a^2}}\Gamma_{v'v}
\end{equation}
\begin{equation}
\label{rate32}
\Gamma_{\bar{3}\bar{2}}=\text{sin}^2\frac{\gamma}{2}~\text{cos}^2\frac{\gamma}{2}~\Gamma_{v'v}[\Delta E_{v'v},~\bm{r}_{--}-\bm{r}_{++}]
\end{equation}
\\\indent The dashed lines in Fig. 2(d) show the calculated relaxation rates relevant to the two experimental values of $E_\textup{VS}$ discussed, including also $B^7$ contribution from SOC mixing with the higher orbital state (see Supplementary Note 3). We use dipole matrix elements as a single free parameter by assuming $|\bm{r}_{-+}|\simeq|\bm{r}_{--}-\bm{r}_{++}|\equiv r$. A least-square fit to the experimental data is performed by fixing the SOC strength to ($\beta_D-\alpha_R$)$\approx 45-60$~m/s (justified by the high electric field $\approx 3\times 10^7$V/m, see Refs.~[\cite{wila,ivchenko}]). The fit then extracts a dipole size $r\approx$1-2~nm for both values of $E_\textup{VS}$.\\\indent
The good agreement between the calculations and experiment, as well as the presence of a hot-spot at the point of degeneracy between Zeeman and valley splitting, provide strong evidence of our ansatz that the spin relaxation is predominantly due to a \textit{new mechanism}: that of mixing with the excited valley states via Rashba/Dresselhaus-like SOC in the presence of interface disorder.\\\indent
Both the splitting at the anticrossing, eq.(\ref{delta_a-B-(110)}), and the intervalley relaxation,
eq.(\ref{valley-relax-Gam0}), depend crucially on the size of the dipole matrix element, $r$,
predicting a fast phonon relaxation of $\approx \, 10^7\, - 10^8\, s^{-1}$
for $r=1-3\, {\rm nm}$, at the hot-spot of Fig.~\ref{T1vsB}(d).
This confirms our core findings that when spin-valley states anticross,
the inter-valley rates are fast for these samples, with
the only available relaxation mechanism being the inter-valley decay.
We point out that these relaxation rates are expected to be sample/material-dependent, given the effect of interface disorder on valley mixing.\\\indent
We now examine the case $N=2$ electrons, and investigate the dependence of the relaxation rate on the magnetic field at a fixed valley splitting ($E_\textup{VS}$=0.58 meV). We note that the energy levels accessible for loading the second electron in the dot, when the $N=1$ spin-down ground state is already occupied, are either the singlet ($S$) or the two lower triplets ($T_-, T_0$), while the higher triplet ($T_+$) would require a spin-flip and is, therefore, not readily accessible [see Fig.~\ref{multi_n}(a)]. In general, for triplet states, the antisymmetry of the two-electron wavefunction requires one electron to occupy a higher energy state. For our  multi-valley QD [Fig.~\ref{device} (f)], this requirement is fulfilled when the two electrons occupy different valley states [see Fig.~\ref{multi_n} (a)]. For low fields, the ground state is $S$ and the triplets have higher energies. This results in excited states (triplets) that extend over two valleys and relax to single-valley ground state (singlet).\\\indent
As the magnetic field is increased, $S$ and $T_-$ undergo an avoided crossing ($B\equiv B_\textup{ST}$), and then $T_-$ becomes the ground state. We adjust the levels of our pulsed readout protocol so that, during the load phase only $S$ and $T_-$ are below the reservoir's Fermi energy, while, during the read phase, the Fermi energy is positioned within the singlet-triplet (ST) energy gap. As a consequence, for $B<B_{ST}$ ($B>B_{ST}$) a $T_-$ ($S$) state would be signalled with a current transient, and relaxation rates can be extracted as for the $N=1$ occupancy. The experimental relaxation rates in Fig.~\ref{multi_n}(b) show a strongly non-monotonic behaviour, approaching an absolute minimum at the anticrossing point ($B_\textup{ST}$=5T). The trend is strikingly symmetric, as can be appreciated when $T_1^{-1}$ is plotted against the detuning energy, as shown in the right inset. This symmetry is reflected in the QD energy spectrum [Fig.~\ref{multi_n}(a)], as far as the detuning $\delta$ is concerned. For $B<B_\textup{ST}$, the ST energy gap decreases with increasing $B$, resulting in slower relaxation rates. By contrast, for $B>B_\textup{ST}$, the ST energy gap increases for increasing field, and so does the relaxation rate.\\\indent
As opposed to the 1-electron case, we note that the 2-electron eigenstates anticrossing leads to a \textit{minimum} in the relaxation rate (\textit{cold-spot}), defined by a splitting at the anticrossing, $\Delta_a^{2e}$, of the same order as that
of eq.(\ref{delta_a-B-(110)}) (see Fig.~\ref{multi_n}(a) and Supplementary Note 3). The occurrence of this minimum does not strictly depend on the nature of the states involved in the decay (spin-like, valley-like, orbital-like or admixtures). It is due to the fact that the avoided crossing takes place between the ground and the first excited state, while for the case $N$=1 it involves the first and the second excited states without affecting the ground state.\\\indent
To model the 2-electron case, we build the wavefunctions for $S$ and $T_-$ from the single-particle states by considering the Coulomb interaction as a perturbing averaged field. The corresponding states are defined as $\left|1\right\rangle_\textup{2e}=\left|v_-,v_-,S\right\rangle$, $\left|2\right\rangle_\textup{2e}=\left|v_-,v_+,T_-\right\rangle$.  Next, the additional perturbation given by SOC leads to renormalized eigenstates which are admixtures of singlet and triplet:
\begin{equation}
\label{1_2e}
\left|\overline{1}\right\rangle_\textup{2e}^{(0)}=\text{sin}\frac{\gamma}{2}\left|1\right\rangle_\textup{2e}-\text{cos}\frac{\gamma}{2}\left|2\right\rangle_\textup{2e}
\end{equation}
\begin{equation}
\label{2_2e}
\left|\overline{2}\right\rangle_\textup{2e}^{(0)}=\text{cos}\frac{\gamma}{2}\left|1\right\rangle_\textup{2e}+\text{sin}\frac{\gamma}{2}\left|2\right\rangle_\textup{2e}
\end{equation}
being similar forms to eq. (\ref{2correct})~(\ref{3correct}). As we show in the Supplementary Note 3, by evaluating the electron-phonon Hamiltonian matrix element for the transition between these states, one finds that it coincides in its form with its 1-electron counterpart for $\left|\overline{3}\right\rangle_\textup{1e}\rightarrow\left|\overline{2}\right\rangle_\textup{1e}$. Therefore, we can conclude that the corresponding relaxation rate, $\Gamma_{\bar{2}\bar{1}}^{2e}$, has the same functional form as those derived in eq. (\ref{rate32}), although the matrix elements for the two cases will be different (see Supplementary Note 3). Dashed lines in Fig.~\ref{multi_n} (b) represent the calculated rates which are fitted to the experimental data similarly to the case $N=1$. Once again, the model convincingly reproduces the main features of the experimental trend, in particular the rates for fields away from the anticrossing together with the symmetry of the characteristics with respect to $B_\textup{ST}$. Further work may be needed to improve the fit in the vicinity of the anticrossing point.  
\\\indent We also measured the relaxation rates for $N=3$ electrons. When the QD occupancy is set at $N$=2, the lower valley is fully occupied and for low $B$-fields the ground state is a singlet. In this condition, the readout protocol is adjusted to probe spin relaxation within the upper valley upon loading/unloading of the third electron. By keeping $E_\textup{VS}=0.58$meV and using the same methodology described before, we measure relaxation rates for the third electron spin state. We find that there is no significant difference between the spin relaxation rates for $N=3$ and $N=1$, as shown in the left inset of Fig.~\ref{multi_n} (b). Two main conclusions can be drawn from this. Firstly, we can infer that the effect of electron-electron interactions on the multi-valley spectrum may be negligible~\cite{wu}, which is plausible. Indeed, for valley 3-electron states, two electrons are just ``spectators'', so that the remaining electron establishes
the same energy level structure as in Fig.~\ref{T1vsB} (c),
and the Coulomb corrections should not affect the valley splitting. Secondly, as we report in Ref. [\cite{ourPRB}], in small QDs for higher occupancies a significantly reduced energy separation between the ground state and the first excited orbital state is observed. This would introduce a non-negligible perturbation on the relaxation if this were affected by the orbital degree of freedom. Hence, the similarities in behaviour in terms of decay rates are a further indication that for our QD the dominant relaxation mechanism resides in the degree of spin-valley admixing, as opposed to the spin-orbit admixing relevant for other semiconductor systems~\cite{hanson}.
\subsection{Valley Splitting Control}
\label{vs ctrl}
\begin{figure*}[]
\includegraphics[scale=0.5]{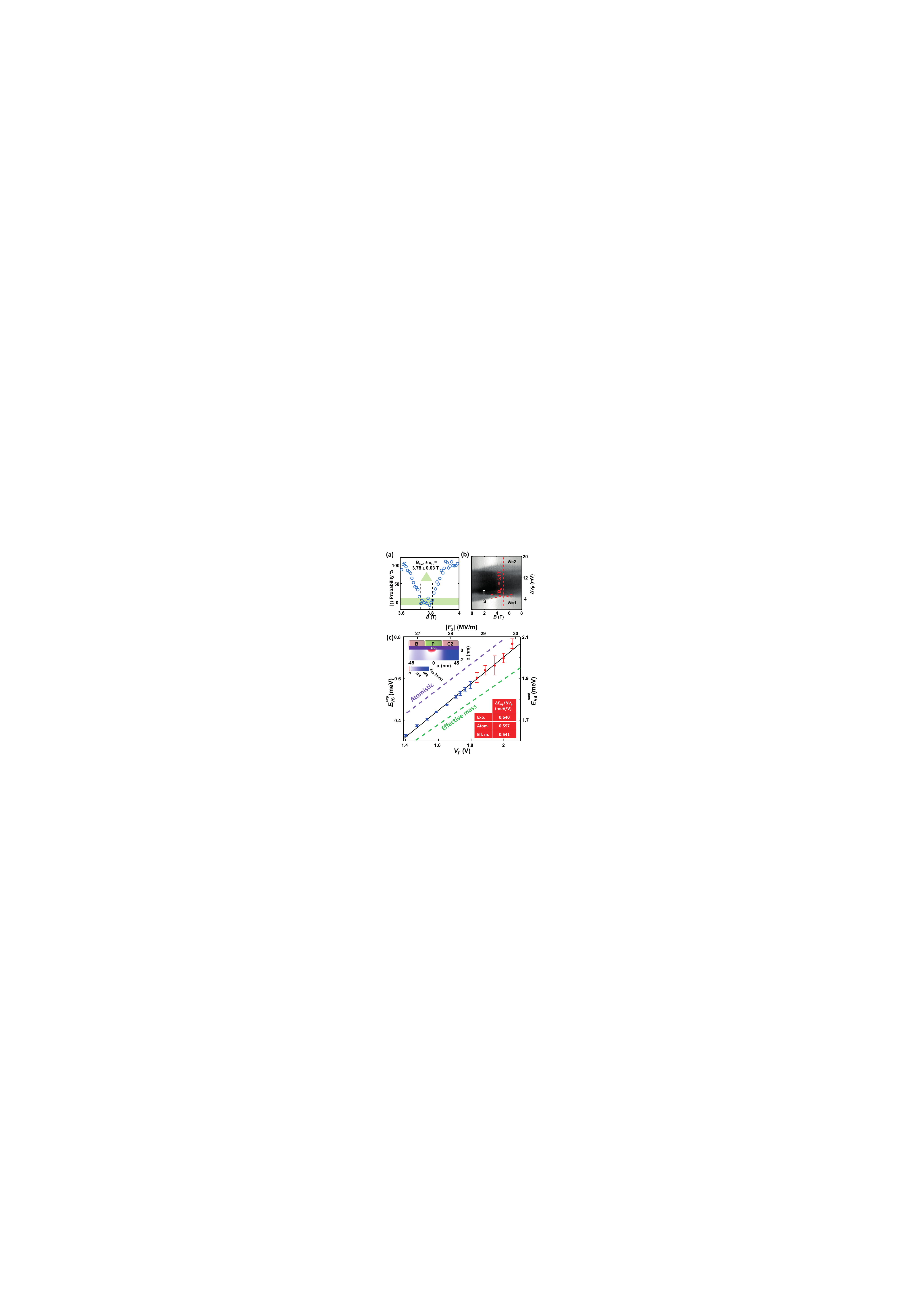}
\caption{Tunability of the valley splitting via gate-voltage control. (a) Spin-up probability as a function of magnetic field for $V_\textup{P}=1.59$V. The occurrence of a hot-spot minimizes the spin-up fraction and allows one to extract $B_\textup{HS}$ (shaded area) and, in turn, $E_\textup{VS}$. (b) Pulsed-voltage magnetospectroscopy showing $dI_\textup{SET}/dV_\textup{P}$ at $N=1\rightarrow 2$ transition. A square pulse of amplitude 16 mV at 287 Hz is applied to gate P. The evolution of the energy difference between the singlet state (light grey) and the triplet state (dark grey) allows one to extract $B_\textup{ST}$ (dashed line) and, in turn, $E_\textup{VS}$. (c) Valley splitting as a function of plunger gate voltage (bottom axis) and modulus of interface vertical electric field in the QD (top axis). Blue and red dots show the valley separation (left axis) measured with hot-spot and magnetospectroscopic techniques, respectively. Error bars indicate standard deviations for the measured values. The linear fit (solid line) indicates a valley tunability of 0.640 meV/V. Dashed lines are calculations performed via atomistic (purple) and effective mass (green) methods (right axis). Bottom inset: Table comparing calculated and experimental tunabilities. Top inset: TCAD simulation of the conduction band profile at $V_\textup{P}$=1.4V in the ($x,z$) plane. Negative energy region in red reveals where the dot is formed.}
\label{fig_vs}
\end{figure*}
We now turn to the experimental demonstration of accurate control of the valley splitting, $E_\textup{VS}$, via electrostatic gating. To determine $E_\textup{VS}$, we use two different experimental approaches. One utilizes the rapid increase in spin relaxation at the hot-spot, and is applicable in the low magnetic field regime. The other is based on magnetospectroscopy, and is relevant for high fields.\\\indent
The first technique stems from the fact that the hot-spot can be reliably detected by monitoring the spin-up probability as a function of magnetic field. In Fig.~\ref{fig_vs}~(a), we show measurements of the spin-up probability performed with the same method as the one used to evaluate spin lifetimes (see Supplementary Note 2). We see that the probability of detecting a spin-up electron decreseas significantly at some magnetic fields. A sudden drop of the spin-up fraction in a narrow range of field identifies the increase in relaxation rate associated with the hot-spot. Given that valley and Zeeman splittings coincide at the hot-spot, one can extract the valley separation as $E_\textup{VS}=g\mu_\textup{B}B_\textup{HS}$, where $B_\textup{HS}$ is defined as the field at which the hot-spot is observed. For varying gate voltage configurations, we scan $B$ in the range 2.8T$<B<$5T, and identify $B_\textup{HS}$ by setting an arbitrary probability threshold [green shaded area in Fig.~\ref{fig_vs}(a)] below which the hot-spot is assumed to occur. The use of this technique is limited to $B<5$T because the lifetime drop at the hot-spot can be therein confidently assessed. At higher fields the relaxation becomes increasingly fast and its enhancement at the hot-spot is indistinguishable within our measurement bandwidth ( $\approx$~10~kHz).\\\indent
In order to evaluate $E_\textup{VS}$ at higher magnetic fields, we use a more conventional magneto-spectroscopic approach, as shown in Fig.~\ref{fig_vs} (b). By employing the same gate-pulsed technique used for the charge stability experiments (see Supplementary Note 1), we focus on the singlet-triplet ground state transition as we load the second electron into the dot (i.e. $N=1\rightarrow2$ transition) in the range 5T$<B<$6.5T. This is clearly identified as the point where the $S$ (light grey feature) and $T_-$ (dark grey feature) states cross. Here, $E_\textup{VS}=g\mu_\textup{B}B_\textup{ST}$, as seen in Fig.~\ref{multi_n}(a).\\\indent
The data points in Fig.~\ref{fig_vs} (c) represent the measured valley separation as a function of $V_\textup{P}$, obtained by means of the aforementioned techniques. The solid line fit shows remarkable consistency between the two sets of data and reveals that $E_\textup{VS}$ depends linearly on the gate voltage over a range of nearly 500 $\mu$eV, with a slope of 640~$\mu$eV/V. In order to keep constant the dot's occupancy and tunnelling rates for different $V_\textup{P}$, a voltage compensation is carried out by tuning gates C1 and B accordingly. We note that we previously reported valley splittings of comparable magnitude (few hundreds of $\mu$eV) in devices realized with the same technology~\cite{ourPRB,Lim2011}. However, to our knowledge, this is the first demonstration of the ability to accurately tune the valley splitting electrostatically in a silicon device.\\\indent
A linear dependence of the valley splitting with respect to the vertical electric field has been predicted for 2DEG systems via effective mass theory~\cite{Saraiva2011,boykin,chutia}. A similar dependence for MOS-based QDs~\cite{sunhee} has also been reported by employing atomistic tight-binding calculations~\cite{klim}. In order to compare our experimental finding with the theoretical predictions, we simulate the vertical electric field ($F_\textup{z}$) in the vicinity of the dot  for the range of gate voltages used in the experiments. We employ the commercial semiconductor software ISE-TCAD~\cite{TCAD} to model the device electrostatic potential, and thereby the electric fields in the nanostructure. For this purpose, TCAD solves the Poisson equation with an approximation of Newton's iterative method~\cite{newton} to obtain convergence at low temperatures.\\\indent
The spatial extent of the dot is identified by regions where the calculated conduction band energy drops below the Fermi level [red area in the top inset of Fig. 4(c)]. Note that our calculations are performed on a three-dimensional geometry identical to the real device with the only free parameter being the amount of offset interface charge. This is adjusted to match the experimental threshold voltage of the device ($V_\textup{th}$=0.625V), as explained in the Supplementary Note 4.\\\indent
The computed variation of interface electric field with gate voltage $V_\textup{P}$ is used to determine the valley splitting according to both the atomistic~\cite{sunhee} and effective mass~\cite{Saraiva2011} predictions. Dashed lines in Fig.~\ref{fig_vs} (c) depict the trends for both approaches, with both exceeding by more than 1 meV the measured values. Despite this offset, the atomistic calculations give a tunability of the valley splitting with gate voltage, $\Delta E_\textup{VS}/\Delta V_\textup{P}$, in good agreement with the experiments. The calculated value of 597 $\mu$eV/V agrees with the measured value to within less than 7\%. The value of 541 $\mu$eV/V calculated using the effective mass approach  reveals a larger deviation ($\approx 15$\%) from the experiments. The presence of an offset in the computed valley splitting may be due to the contribution of surface roughness that is not accounted for in the models, and is thought to be responsible for a global reduction of $E_\textup{VS}$~\cite{Saraiva2009,chutia,Culcer2010,friesen,andosolo,fec}. We emphasize, however, that the gate tunability would remain robust against this effect, which is not dependent on electric field.
\section{Discussion}
\label{conclusions}
In this work, we have shown that the valley splitting in a silicon device can be electrostatically controlled by simple tuning of the gate bias. We used this valley splitting control, together with spin relaxation measurements, to explore the interplay between spin and valley levels in a few-electron quantum dot. \\\indent
The relaxation rates for a one-electron system exhibit a dramatic hot-spot enhancement when the spin Zeeman energy equals the valley splitting, while for a two-electron system the rates reach a minimum at this condition. We found that the known mechanisms for spin relaxation, such as the admixing of spin and $p$ orbital states, were unable to explain the key features of the experimental lifetime data, and so introduced a novel approach based on admixing of valley and spin eigenstates. Our theory, which showed good agreement with experiment, implies that spin relaxation via phonon emission due to spin-orbit coupling can occur in realistic quantum dot systems, most likely due to interface disorder.\\\indent
Our results show that by electrical tuning of the valley splitting in silicon quantum dots, it is possible to ensure the long lifetimes ($T_1>1$~s) required for robust spin qubit operation. Despite this, the excited valley state will generally be lower than orbital states in small quantum dots, placing an ultimate limit on the lifetimes accessible in very small dots, due to the spin-valley mixing described above. \\\indent
Electrical manipulation of the valley states is also a fundamental requirement to perform coherent valley operations. However, the experimental relaxation rate at the observed hot-spot was found to be fast ($T_1^{-1}>1$~kHz) for our devices, implying a fast inter-valley relaxation rate.\\\indent
Finally, in the context of realizing scalable quantum computers, these results allow us to address questions of device uniformity and reproducibility with greater optimism. Indeed, our work suggests that issues related to the wide variability of the valley splitting observed in silicon nanostructures to date can shift from the elusive atomic level (surface roughness, strain, interface disorder) to the more accessible device level, where gate geometry and electrostatic confinement can be engineered to ensure robust qubit systems.

\section{Methods}
\label{methods}
\subsection{Device Fabrication}
The samples fabricated for these experiments are silicon MOS planar structures. The high purity, near intrinsic, natural isotope silicon substrate has n+ ohmic regions for source/drain contacts defined via phosphorous diffusion. High quality SiO$_2$ gate oxide is 8 nm thick and is grown by dry oxidation at 800$\degree$C. The gates are defined by electron-beam lithography, Al thermal evaporation and oxidation. Three layers of Al/Al$_2$O$_3$ are stacked and used to selectively form a 2DEG at the Si/SiO$_2$ interface and provide quantum confinement in all the three dimensions.\\\indent
\subsection{Measurement System}
Measurements are carried out in a dilution refrigerator with a base temperature T$_b\approx$~40mK. Flexible coaxial lines fitted with low-temperature low-pass filters connect the device with the room-temperature electronics. In order to reduce pick-up noise, the gates are biased via battery powered and opto-isolated voltage sources. The SET current is amplified by a room-temperature transimpedance amplifier and measured via a fast digitizing oscilloscope and a lock-in amplifier for the single-shot and energy spectrum experiments, respectively. Gate voltage pulses are produced by an arbitrary wave-function generator and combined with a DC offset via a room-temperature bias-tee.

\section*{ACKNOWLEDGMENTS}

The authors thank D. Culcer for insightful discussions, and F. Hudson and D. Barber for technical support. This work was supported by the Australian National Fabrication Facility, the Australian Research Council (under contract CE110001027), and by the U.S. Army Research Office (under contract W911NF-13-1-0024). The use of nanoHUB.org computational resources operated by the Network for Computational Nanotechnology funded by the US National Science Foundation under grant EEC-0228390 is gratefully acknowledged.
\section*{FINANCIAL INTERESTS}
The authors declare no competing financial interests.
\section*{AUTHOR CONTRIBUTIONS}
C.H.Y. carried out the measurements. N.S.L. designed and fabricated the device. C.H.Y., A.R. analyzed the data. C.H.Y., A.R., S.L., G.K., A.M., A.S.D. discussed the results. R.R., C.T. modelled the relaxation rates. F.M. simulated the electric field profiles. A.S.D. conceived and planned the project. A.R. wrote the manuscript with input from all coauthors.

\newpage
\section*{\\[2in]SUPPLEMENTARY INFORMATION}
\newpage
\begin{suppfigure*}[]
\includegraphics[scale=0.5]{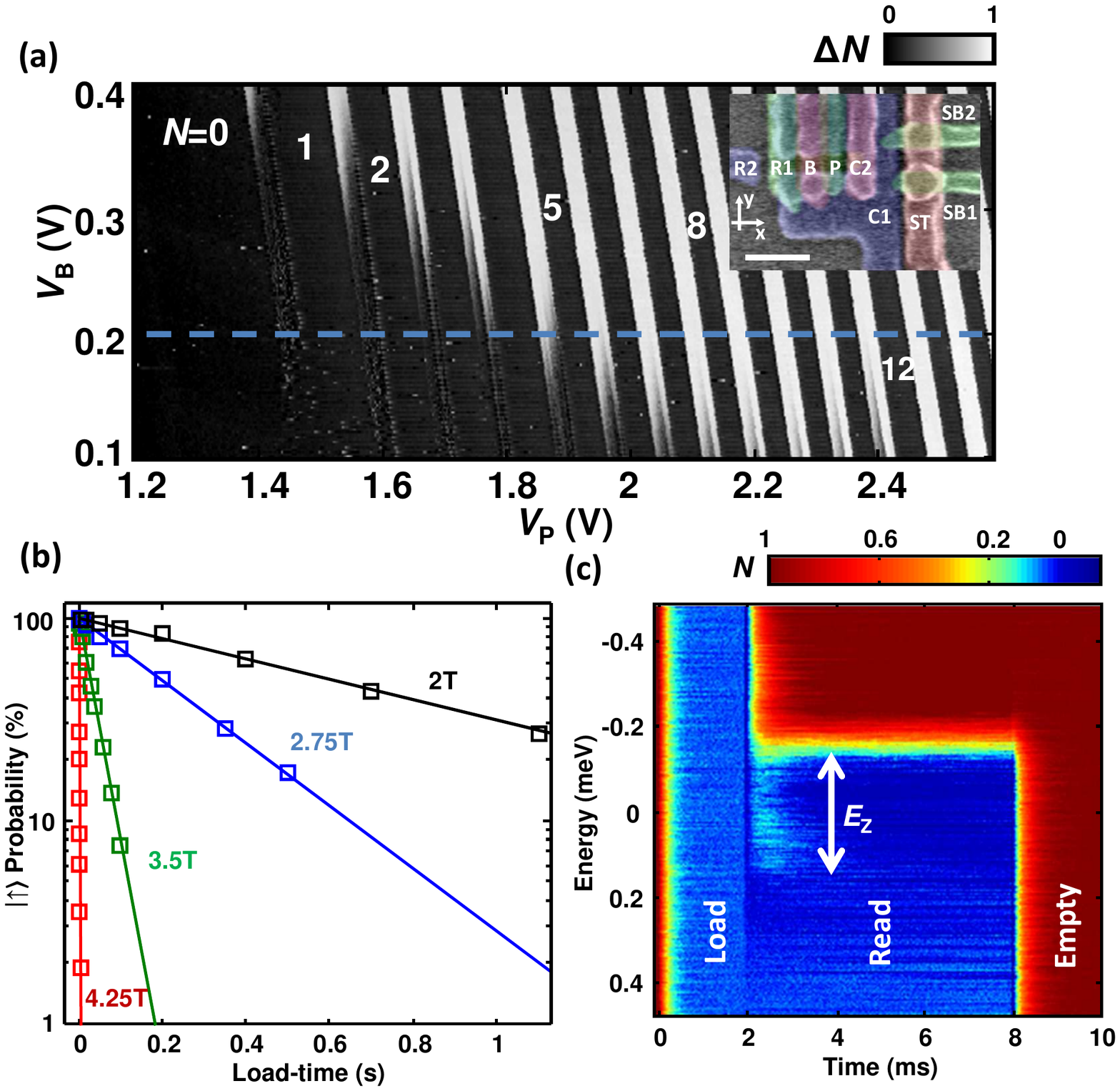}
\caption{Charge and spin detection. (a) Pulsed-voltage charge stability diagram with an applied square pulse of 32 mV peak-to-peak at 287 Hz. Grey-scale indicates the excess electron occupancy in the dot ($\Delta N$) for each charge addition. Inset: False-color scanning electron micrograph of a device identical to the one used for experiments. Color code identifies different layers in the gate stack. Scale bar length is 100 nm. (b) Exponential decays of the spin-up probabilities at different magnetic fields. Solid lines are single-exponential fits of the measured data (squared points) leading to the evaluation of T$_1$ times. (c) QD occupancy during a 3-level pulse with varying read-phase level. The energy axis shows the dot's electrochemical potential during the read phase. Each trace is averaged over 256 single-shots. Applied field $B=3$ T. Zeeman splitting can be evaluated from the current transient energy range.}
\end{suppfigure*}


\newpage 
\subsection*{Supplementary Note 1}
The addition spectrum of the QD is characterized via a technique widely used for GaAs-based systems, which employs both charge detection and gate pulsing$^{30}$. A train of square voltage pulses is applied to gate P, in addition to its dc voltage, shifting the energy levels of the dot up and down. This pulse train modulates the SET sensor current with the same frequency of the pulse due to cross-capacitance effects. The resulting current at this frequency is measured with a lock-in amplifier. Full details of the measurement set-up are reported elsewhere$^{41}$. Supplementary Figure S1 (a) shows the measured stability map as a function of plunger gate voltage, $V_\textup{P}$, and barrier gate voltage, $V_\textup{B}$ (see inset). Current peaks in the lock-in signal indicate the occurrence of charge transitions due to tunnelling between the QD and the reservoir. We have converted this signal into an average change of occupancy$^{41}$, $\Delta N$, which is shown in the grey-scale plot. For each transition, we can also probe the first orbital excited state whenever a slow decrease in the detection signal is followed by a steep rise$^{41}$. We observe charge additions for the first fourteen electrons in the dot. In order to translate the voltage dependence of these features into a spectroscopy of the dot's charging energy ($E_\textup{C}$), we determine the voltage to energy conversion factor, $\alpha$, for gate P  via temperature dependence measurements and we find $\alpha\approx$~0.16 eV/V. The dashed horizontal line indicates that the addition spectrum shown in the main article has been obtained for $V_\textup{B}$=0.2V.
\section*{Supplementary Note 2}
Here, we describe the procedure utilized to measure spin lifetimes for varying $B$. The probability of observing a spin-up electron, $P_{up}$, decreases when the loading time, $t_{load}$, is increased. A threshold detection method$^{24}$ is used to determine the fraction of shots that contribute to the spin-up count for each $t_{load}$ at different $B$. Supplementary Figure S1 (b) shows that the probability is well fitted by a single exponential time decay, $P_{up}\propto e^{-t_{load}/T_1}$, from which $T_1(B)$ is evaluated.\\\indent
By modifying the pulse sequence and recording readout traces for different voltage levels of the read phase, it is possible to extract the single-particle Zeeman splitting$^{24}$. Indeed, by stepping the read voltage from a level where both spin states are lifted above the Fermi energy of the reservoir to a level where both spin states are pushed below the reservoir's energy, the separation between the two spin states can be evaluated. Supplementary Figure S1 (c) shows the QD occupancy averaged over 256 single-shot traces for varying read levels at $B$=3T (note that the energy scale on the y-axis is the effective shift of the dot's potential). As explained in the main article, the detection of a spin-up electron would be signalled by a transient in the detector's current. This transient is seen to extend for about 350$\mu$eV, as expected from the Zeeman equation, $E_\textup{Z}=g\mu_\textup{B}B$, by assuming $g=2$. The field dependence of the measured $E_\textup{Z}(B)$ also shows good agreement with the Zeeman equation. This proves that we can reliably load/unload electrons onto spin-split single-particle states.\\
\section*{Supplementary Note 3}

We consider a small quantum dot (QD)  defined electrostatically in a MOS Si/SiO$_2$ heterostructure,
as described in the main text.
For a QD occupied by a single electron the first excited orbital state is some 8 meV above the
ground state and it only weakly influences the ground state shell (see however below).
Due to strong electric field at the heterostructure,
the 6-fold degeneracy of the conduction band electrons is lifted:
only two lowest energy valleys remain relevant, at momentums
$\bm{k} \approx k_0\, [\hat{z}, -\hat{z}]$, with $k_0  \simeq  0.85\, \frac{2\pi}{a_0}$;
the remaining two-fold degeneracy is lifted via the sharp interface
potential, leading to a valley splitting, $E_\textup{VS}$, of the order of few hundred $\mu{\rm eV}$,
and linearly proportional to the electric field at the interface,
$E_\textup{VS} \simeq \alpha_E \langle E_{\rm interface}\rangle$.
This forms the ground state shell  of the two lowest valley states,  $v_1$ and $v_2$;
(these approach states of definite parity, $v_{1,2} \to v_{-,+}$ in strong field, as per notation in the main article).

In the effective mass approximation the valley wave functions are generally
written as
\begin{eqnarray}
&& |v_1\rangle = \sum_{i=\pm z} \Phi_{v1}^i(\bm{r}) \alpha_i^{v_1} u_i(\bm{r})\, e^{i k_i z}
\label{v1-1e}\\
&& |v_2\rangle = \sum_{i=\pm z} \Phi_{v2}^i(\bm{r}) \alpha_i^{v_2} u_i(\bm{r})\, e^{i k_i z}
\label{v2-1e}  ,
\end{eqnarray}
where $\psi_i(\bm{r}) \equiv u_i(\bm{r})\, e^{i k_i z}$ are the Kohn-Luttinger valley functions,
with the periodic  part $u_i(\bm{r})$, and $k_i$ ($i=\pm z$) is the position of the valley minima.
%
Here, $\Phi_{v_j}^i(\bm{r})$ are the envelope functions corresponding to the
orbital ground state ($s$-like).
In the ideal QD (ideal interface, symmetric circular dot, etc.) the envelopes are separable, e.g.:
$\Phi_0^{+z}(\bm{r}) = F_0(x,y)\, f_0(z)$;
generally, the index of the state, $v_1$ or $v_2$, distinguishes different envelopes,
originating from valley-orbit coupling in the presence of a non-ideal interface,
also separability may be lost.
The relations: $\Phi_{v_j}^{+z} = \Phi_{v_j}^{-z}$ and $|\alpha_{+z}^{v_j}| = |\alpha_{-z}^{v_j}|$
are maintained for each state by time reversal.
For the states of definite parity above, the valley populations are:
$\alpha_{+z}^{-} = -\alpha_{-z}^{-} = \frac{1}{\sqrt{2}}$,
$\alpha_{+z}^{+} = \alpha_{-z}^{+} = \frac{1}{\sqrt{2}}$.

By switching on a magnetic field, the states are Zeeman split and
one gets four lowest spin-valley states denoted as
$|1\rangle\! = |v_1,\!\!\downarrow\rangle$,
$|2\rangle\! = |v_1,\!\!\uparrow\rangle$,
$|3\rangle\! = |v_2,\!\!\downarrow\rangle$,
$|4\rangle\! = |v_2,\!\!\uparrow\rangle$.
The corresponding (unperturbed) energies are  $E_1 = - E_\textup{Z}/2$,
$E_2 = + E_\textup{Z}/2$, $E_3 = E_\textup{VS} - E_\textup{Z}/2$, $E_4 = E_\textup{VS} + E_\textup{Z}/2$,
where levels 2 and 3 cross at $E_\textup{Z}=E_\textup{VS}$,
and $E_\textup{Z} = g_{Si} \mu_B B$ is the Zeeman splitting.
Since electron-phonon interaction does not flip spin, phonon relaxation
between $|2\rangle$ and $|1\rangle$ is forbidden, while $|3\rangle$ to $|1\rangle$ is allowed as
a pure valley (non-spin flip) relaxation.
It is the mixing  between valley states $|2\rangle$ and $|3\rangle$ due to spin-orbit coupling (SOC)
that causes  a  renormalized  state  $|\bar{2}\rangle$ to have a spin-down  component,
that allows phonon transition  $|\bar{2}\rangle \to |1\rangle$ to take place.
The SOC interaction  forms  an anticrossing  point for the states
$|\bar{2}\rangle$, $|\bar{3}\rangle$,
with a splitting  $\Delta_a$, where the spin flip is maximal
and so is the $|\bar{2}\rangle \to |1\rangle$ relaxation, reaching the order of pure phonon-valley
relaxation, which is observed as a relaxation ``hot-spot'' (see Fig. 2 (c,d) of main text).

Both the anticipated non-zero splitting at the anticrossing  of the two valley states,
and the phonon relaxation mechanism
requires non-zero valley dipole matrix elements:
$\bm{r}_{12} \equiv  \langle v_1| \bm{r}| v_2\rangle$,
$\bm{r}_{11} \equiv  \langle v_1| \bm{r}| v_1\rangle$, and
$\bm{r}_{22} \equiv  \langle v_2| \bm{r}| v_2\rangle$.
In the ideal case, however, they
will take a non-zero value only from Umklapp (intervalley) contributions,
which are highly suppressed in a QD.
While a complete theory of the valley wave functions in the presence of a roughness/steps
is still under development, some gross effects of the experiment can be explained
just by {\it postulating} non-zero dipole matrix elements. 
The intravalley contribution is generally given by (it is zero in the ideal case):
\begin{eqnarray}
&& \bm{r}_{12} =\langle v_1| \bm{r} | v_2\rangle
\nonumber\\
&& \qquad \simeq  \sum_{i=\pm z} \alpha_i^{v_1*} \alpha_i^{v_2}
\int d\bm{r} \, \bm{r} \, \Phi_{v1}^{i*}(\bm{r}) \Phi_{v2}^{i*}(\bm{r})\, u_i^{*}(\bm{r}) u_i(\bm{r})
\label{dipole-me} ,
\end{eqnarray}
and similarly for $\bm{r}_{11}$, $\bm{r}_{22}$.
Numerical fit to the experimental data presented in this paper
consistently reveal values of $|\bm{r}_{12}|$ the order of few nm
(these may depend on the applied electric field at the interface,
and also on the QD occupancy).  

To calculate the splitting at anticrossing of the spin-valley states,
$|2\rangle\! = |v_1,\!\!\uparrow\rangle$, and
$|3\rangle\! = |v_2,\!\!\downarrow\rangle$,
we use the spin-orbit coupling (SOC).
In two dimensions  the well known SOC Hamiltonian reads:
$H_\textup{SO} = H_\textup{D} + H_\textup{R}$,
with the Dresselhaus and Rashba terms,
related to conduction band  electrons, confined in a $[001]$  2DEG:
\begin{equation}
H_D = \beta_D (-\sigma_x\, P_x + \sigma_y\, P_y), \ \ H_R = \alpha_R (\sigma_x\, P_y + \sigma_y\, P_x)
\label{SOC} ;
\end{equation}
here $\sigma_x$, $\sigma_y$ are the Pauli matrices along the principal crystal axes,
and $\bm{P} = \bm{p} + e \bm{A}(\bm{r})$  is  the generalized momentum.
Related to current experiment, we consider an in-plane magnetic field, which is parallel
to the  $(110)$-direction, and its
vector potential in the symmetric gauge  reads:
$\bm{A}(\bm{r}) = \frac{B}{2\sqrt{2}}\,(z,-z,-x+y)$.
Using the eigenoperators $\sigma_{\pm 45}\equiv -(\sigma_x \pm \sigma_y)$,
($\sigma_{+45} |\uparrow,\downarrow\rangle = (\pm)|\uparrow,\downarrow\rangle$),
and Hamiltonian commutation  relations
one can express the perturbation matrix
via dipole matrix elements, e.g.:
\begin{equation}
V_{23} = \langle v_1,\!\uparrow| H_{SO}|v_2,\!\downarrow\rangle
= \frac{i\, m_t\, E_{VS}}{\sqrt{2} \hbar} (x_{12} + y_{12}) (\beta_D - \alpha_R)
\label{B-(110)},
\end{equation}
while $V_{12}, V_{13} \approx 0$.
Estimation for the splitting gives:
$\Delta_a = 2 |V_{23}| \approx r\times 0.7\, 10^{-4}\, E_\textup{VS}$,
where, we have used transverse mass $m_t \simeq 0.198 m_e$,    
and $x_{12} = y_{12} \equiv r$ is the dipole size ($z_{12} \approx 0$).
We assume a SOC strength $\beta_D - \alpha_R \approx 15\, {\rm m/s}$
for an interface field of $10^7\, {\rm V/m}$
(implying   
it may reach $3-4$ higher values in the current experiment).
%
Note, that in general, $\Delta_a$  depends  linearly on
the valley splitting and the dipole matrix element.

To calculate the phonon relaxation rate between the lowest spin-valley states
we take into account the
hybridization of the levels $|2\rangle$ and $|3\rangle$ due to SOC
[see Fig.~2(c)].
The relevant matrix elements
\begin{eqnarray}
&& \langle \bar{2} | H_{\rm e-ph} | 1\rangle = -\cos{\frac{\gamma}{2}}\, \langle v_2 |H_{\rm e-ph} | v_1 \rangle
\label{relaxat-me21}
\\
&& \langle \bar{3} | H_{\rm e-ph} | 1\rangle = \sin{\frac{\gamma}{2}}\, \langle v_2 |H_{\rm e-ph} | v_1 \rangle
\label{relaxat-me31}
\\
&& \langle \bar{3} | H_{\rm e-ph} | \bar{2}\rangle = \frac{1}{2}\sin{\gamma}\,
\left[  \langle v_1 |H_{\rm e-ph} | v_1 \rangle - \langle v_2 |H_{\rm e-ph} | v_2 \rangle\right]
\label{relaxat-me32}
\end{eqnarray}
are expressed via the admixture of the state $|3\rangle = |v_2,\!\downarrow \rangle$,
to  the states $|\bar{2}\rangle$,  $|\bar{3}\rangle$  (Eqs.~(2),(3) of the main text)
and via the phonon transition matrix element between the valley states.

 Calculation of the valley phonon relaxation rate is performed via the
electron-phonon deformation potential interaction~\cite{Cardona}
in the approximation when only intravalley contributions are
taken into account, and using the non-zero dipole matrix element,
Eq.(\ref{dipole-me}). %
For the relaxation  below the anticrossing ($|\bar{2}\rangle \to |1\rangle$)
we obtain
$\Gamma_{\bar{2}1}^{(\sigma)} = \cos^2{\frac{\gamma}{2}}\, \Gamma_{v'v}^{(\sigma)}$.
Above anticrossing, the experimentally observable transitions are
$|\bar{3}\rangle \to |1\rangle$ and $|\bar{3}\rangle \to |\bar{2}\rangle$, see Fig.~2 (c) of main text.
(In the second case, a subsequent decay of $|\bar{2}\rangle \to |1\rangle$  above anticrossing is
purely valley and therefore very fast).
The rates are:
$\Gamma_{\bar{3}1}^{(\sigma)} = \sin^2{\frac{\gamma}{2}}\, \Gamma_{v'v}^{(\sigma)}$
and
$\Gamma_{\bar{3}\bar{2}}^{(\sigma)} = \cos^2{\frac{\gamma}{2}}\,\sin^2{\frac{\gamma}{2}}\, \Gamma_{v'v}^{(\sigma)}$,
respectively.
In all three cases above,
the valley phonon relaxation rate  $\Gamma_{v'v}^{(\sigma)}$
is of the same functional form
(we used dipole approximation):
%
\begin{eqnarray}
&& \Gamma_{v'v}^{(\sigma)}\left[\Delta E_{v'v},\bm{r}_{12}\right]  = \frac{1}{4\pi \rho \hbar}\, \frac{\omega_{v'v}^5}{v_{\sigma}^7}
\, \int_{-1}^1 dx \, D(x,\bm{r}_{12}) \, \Xi^{(\sigma)}(x) \qquad
\label{valley-relax-Gam}\\
&& D(x,\bm{r}_{12})  \equiv \frac{x_{12}^2 + y_{12}^2 }{2} (1-x^2) + z_{12}^2\,x^2, \ \ x\equiv \cos(\theta)
\label{valley-relax-D}  ,
\end{eqnarray}
and depends on the actual transition energy $\Delta E_{v'v}$;    
here $D(x,\bm{r}_{12})$ depends quadratically on the dipole components $\bm{r}_{12}$.
Note, that for the transition $|\bar{3}\rangle \to |\bar{2}\rangle$  one have to substitute
$\bm{r}_{12}$ by the difference dipole matrix element $\bm{r}_{11}-\bm{r}_{22}$.
The integrand, that is proportional to the deformation potential constants, $\Xi_d, \Xi_u$, is:
$\Xi^{(l)}(x) = \Xi_d^2 + 2 \Xi_d \Xi_u x^2 + \Xi_u^2 x^4$ for longitudinal phonons, and
$\Xi^{(t2)}(x) = \Xi_u^2 x^2 (1-x^2)$ for transverse phonons.
\begin{suppfigure*}[t]
\includegraphics[scale=0.5]{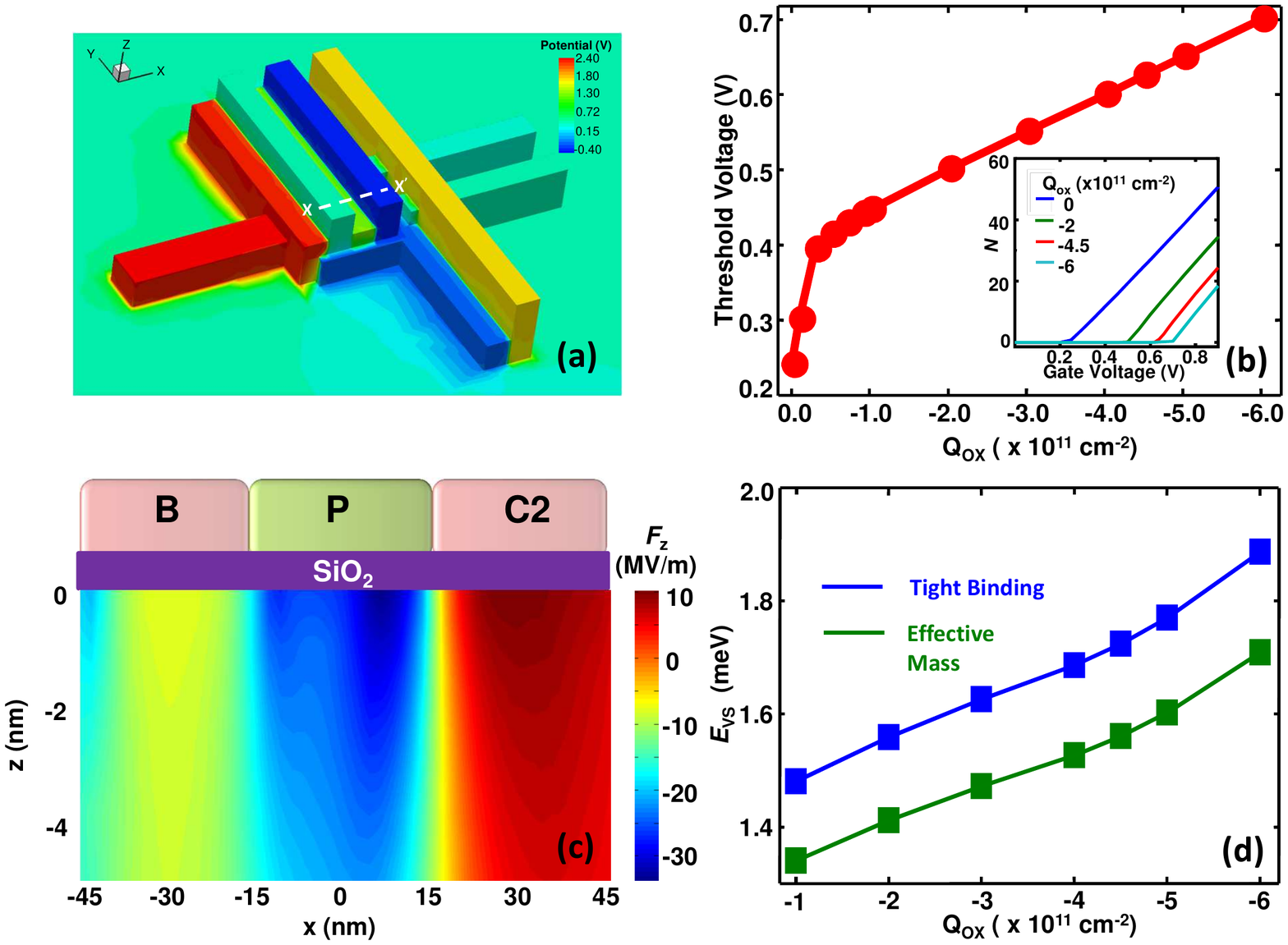}
\caption{TCAD simulations. (a) Three-dimensional device layout used for the simulations. Color scale represents the electrostatic potential at each gate for one iteration. (b) Calculated threshold voltage as a function of interface charge. The experimental value (0.625V) is obtained at $Q_\textup{ox}=-4.5\times 10^{11}$~cm$^{-2}$. Inset: integrated electron density in the dot vs the voltage applied simultaneously to all gates for different $Q_\textup{ox}$. Threshold voltage dependence on $Q_\textup{ox}$ can be extracted (main graph). (c) Two-dimensional electric field profile at $V_\textup{P}$=1.4V. (d) Valley splitting at $V_\textup{P}$=1.4V as a function of interface charge calculated with the atomistic (blue) and effective mass (green) predictions.}
\label{fig_s2}
\end{suppfigure*}
Using Eq.(\ref{valley-relax-Gam}), one can see that the dependence of the relaxation
on magnetic field is $\sim B^5$  far below anticrossing,
since the spin admixture coefficient $\cos^2{\frac{\gamma}{2}} \sim \frac{\Delta_a^2}{4 (E_\textup{VS} - E_\textup{Z})^2}$
is only weakly $B$-dependent for $E_\textup{Z} \ll E_\textup{VS}$.
At anticrossing, the mixing between spin-up and spin-down states is maximal,
and relaxation reaches a fast (pure valley) transition rate;
e.g., for $B=2.8$T, 
$\Gamma_{\bar{2}1} \approx \, 10^7\, - 10^8\, s^{-1}$
for $r=1-3\, {\rm nm}$.
Numerical fits to the experimental data show that
the intervalley SOC mixing mechanism alone,
is not enough
to describe the actual $B$-dependence  above anticrossing (except close vicinity of the hot-spot).
An additional mechanism of virtual excitation of the first orbital $2p$-state
with overall $B^7$ dependence becomes important at high magnetic fields (see below).

{\it Spin-valley relaxation mechanism in the 2 electron-case.}\\
In considering the two-electron case
one replaces  the Coulomb interaction by a mean field:
$U(\bm{r}_1,\bm{r}_2) \approx \tilde{U}(\bm{r}_1) + \tilde{U}(\bm{r}_2)$,
and the 2-electron wave functions are constructed from the corresponding single-particle states,
$\varphi_i(\bm{r}) \approx |v_i\rangle,\, i=1,2$, Eqs. (\ref{v1-1e}),(\ref{v2-1e}).
[The single-particle states in the 2-electron case should include corrections due to the
mean field potential, neglected in zeroth-order approximation].
The corresponding spin-valley states  are the spin-singlet and spin-triplet:
%
\begin{eqnarray}
&& |v_1',S \rangle = \varphi_1(\bm{r}_1) \varphi_1(\bm{r}_2)\otimes \frac{1}{\sqrt{2}}
\left[ |\uparrow_1 \downarrow_2 \rangle - |\downarrow_1\uparrow_2 \rangle  \right]
\label{2e-Singlet}\\
&& |v_2',{T_-} \rangle = \frac{1}{\sqrt{2}}\,
\left[ \varphi_1(\bm{r}_1) \varphi_2(\bm{r}_2) - \varphi_1(\bm{r}_2) \varphi_2(\bm{r}_1)  \right]  \otimes
|\downarrow_1 \downarrow_2 \rangle \qquad
\label{2e-Triplet}  .
\end{eqnarray}
For the energy difference in the 2-electron case, one have to take account of the
Coulomb interaction in the two configurations:
$\Delta E_\textup{TS} = E_\textup{T} - E_\textup{S} = E_\textup{VS} + \langle U\rangle_\textup{T} - \langle U\rangle_\textup{S}$.
Since the valley states are $s$-like (and using mean field approximation)
one can show that
the 2-electron valley splitting  coincides  with the 1-electron case:
$\Delta E_\textup{TS} \simeq E_\textup{VS}$.

Similar to the 1-electron case, one calculates the splitting at anticrossing
for the levels $|1\rangle_{2e} =|v_1',S\rangle$ and $|2\rangle_{2e} =|v_2',{T_-}\rangle$.
For magnetic field along the $(110)$-direction the non-diagonal matrix element coincides with
the single-particle one, Eq.(\ref{B-(110)}):
\begin{equation}
V_{12}= \langle 1| H_{SO} |2\rangle = \frac{i m_t E_\textup{VS} (\beta_D - \alpha_R)}{\sqrt{2}\hbar}
\langle \varphi_1(\bm{r}_1)| (x_1 + y_1) |\varphi_2(\bm{r}_1)\rangle
\label{split-(110)-2e} ,
\end{equation}
and the splitting is the same as well: $\Delta_a^{2e} = 2|V_{12}| = \Delta_a$.
Thus, the hybridized states (in the presence of SOC),
are admixtures of $|1\rangle_{2e}$, $|2\rangle_{2e}$, with the
same admixture coefficients $\sin{\frac{\gamma}{2}}$, $\cos{\frac{\gamma}{2}}$, as for the single-electron case.
The  electron-phonon  matrix element can be calculated:
$\langle \bar{2} | H_{\rm e-ph} | \bar{1}\rangle = \frac{1}{2}\sin{\gamma}\,
\left[  \langle v_1',S |H_{\rm e-ph} | v_1',S \rangle - \langle v_2',T_{-} |H_{\rm e-ph} | v_2',T_{-} \rangle\right]$
and shown to coincide  with the 1-electron matrix element of the $\bar{3} \to \bar{2}$ transition:
$\langle \bar{2} | H_{\rm e-ph} | \bar{1}\rangle = \langle \bar{3} | H_{\rm e-ph} | \bar{2}\rangle$.
Therefore, the corresponding rates also coincide:
$\Gamma_{\bar{2}\bar{1}}^{(\sigma)} =\Gamma_{\bar{3}\bar{2}}^{(\sigma)} = \cos^2{\frac{\gamma}{2}}\, \sin^2{\frac{\gamma}{2}}\
\Gamma_{v'v}^{(\sigma)}(\left[\Delta E_{v'v},\bm{r}_{11}-\bm{r}_{22} \right]$,
as the energy difference is the same for both transitions.
Numerical fit to the data just above anticrossing in the 1-electron case
for $E_\textup{VS}=0.33$ meV, and a fit to the data in the 2-electron case
for $E_\textup{VS}=0.58$ meV (see Fig. 2(d) and Fig. 3(b) of the main text)
are consistent with dipole matrix element
$\bm{r}_{11}-\bm{r}_{22}$ of few nm.
Therefore, the experiment inevitably shows that the difference dipole matrix element is
non-zero,
$\Delta r_{12} \equiv \bm{r}_{11}-\bm{r}_{22} \neq 0$.

{\it Including $B^7$-corrections due to virtual transitions to orbital 2p-states. 1-electron case}\\
In the one electron case we consider orbital excitations
above the ground state $|v_1\rangle$, split by $\Delta \simeq 8\,\mbox{meV}$.
For an in-plane magnetic field with weak magnetic confinement,
we have $\omega_{x,y} \gg \omega_c$ and these states
are separable$^{51}$:
$|m_1\rangle = |0_x\, 1_y\rangle ,\ \ |m_2\rangle = |1_x\, 0_y\rangle$.
The corresponding non-zero dipole matrix elements are
$\langle 0 |y |m_1\rangle \equiv y_{0m_1} = 
\langle 0 |x |m_2\rangle \equiv x_{0m_2} = \sqrt{\frac{\hbar^2}{2 m_t \Delta}}$.
SOC  makes the qubit states, $|1\rangle$, $|2\rangle$, to mix with these
upper orbital states. Calculation in the first order of PT
is performed (similarly to Ref.[38]), while we neglect small corrections
due to hybridization of the states $|2\rangle$, $|3\rangle$.
The resulting rate has the $B^7$ dependence
on the magnetic field (expected for a $1s \to 2p$ virtual transition):
\begin{equation}
\Gamma_{2p}^{1e} = \left[ \frac{E_Z^2 \Delta^2}{(\Delta^2-E_Z^2)^2} \right]
\frac{m_t^2 (\beta_D - \alpha_R)^2 E_{v'v}^5}{4\pi \rho \hbar^8}(TA+LA)\,
\left[ y_{0m_1}^4 + x_{0m_1}^4 \right]
\label{2p-1e-case} ;
\end{equation}
here
$(TA+LA) \equiv \int_{-1}^1 dx\, \Xi^{(\sigma)}(x) (1-x^2)/v_{\sigma}^7$
accounts for contribution of transverse and longitudinal acoustic phonons.
Numerical fit to the 1-electron experimental data
for $E_\textup{VS} = 2.8\,T$ shows that this relaxation mechanism
has to be taken in to account in order to describe the data above anticrossing.
The fit is consistent with $1s-2p$ dipole matrix elements about twice bigger
than for a parabolic QD, of the order of $\approx 10\, {\rm nm}$.

{\it 2-electron case $\Delta E^7$ corrections}\\
An analogous calculation in the 2-electron case was performed as well.
We examined virtual excitations to the whole set of excited states
that includes the triplet valley states, the upper singlet valley state,
and the singlet and triplet states, involving orbital single-electron states.
We found that the only relevant state
(capable to produce a relaxation rate correction of $\Delta E^7$ behavior)
is the valley-spin-orbital state:
$|m_1'',{T} \rangle =
\frac{1}{\sqrt{2}}\,
\left[ \varphi_2(\bm{r}_1) \varphi_{m1}(\bm{r}_2) - \varphi_2(\bm{r}_2) \varphi_{m1}(\bm{r}_1)  \right]
\otimes |{T_{-}} \rangle$,
where one of the electrons is occupying $|v_2\rangle$ state, while the other is orbitally excited.
%
We have derived a contribution to the relaxation rate,
$\Gamma_{2p}^{2e} \propto \sim E^7$, similar to the 1-electron case.
%
%
%
However, including this contribution  does not help to fit better
the 2-electron data for reasonable parameters.

{\it Best fit parameters}\\
For the best fit to the 1-electron and 2-electron experimental data
we use:\\
(1) SOC strength $\beta_D - \alpha_R \approx 45-60\, {\rm m/s}$, which
is justified, since the electric field in the experiment is $\approx 3\times 10^7\, {\rm V/m}$.\\
(2) We take (for each $E_\textup{VS}$) that $|\bm{r}_{12}| \simeq |\Delta\bm{r}_{12}| \equiv |\bm{r}_{11}-\bm{r}_{22}|$
(justified by common sense and preliminary calculations).\\
(3) Splitting to the first orbital excited state, $\Delta = 8\, {\rm meV}$ (1-electron case)
and $\Delta_2 = 4\, {\rm meV}$ (2-electron case) taken from experimental values.\\
(4) In the 1-electron fit, the $1s-2p$ dipole matrix elements are $x_{0m_2} = y_{0m_1} \simeq 1.8\, \sqrt{\frac{\hbar^2}{2 m_t \Delta}}$
(a plausible assumption for a non-parabolic dot),
and in the 2-electron case we take the same values for these matrix elements.
Least-square fit to the data then reveals the
valley dipole matrix elements for each set of data, as follows:
$r_{28}=1.1\, {\rm nm}$ ($E_\textup{VS}=0.33$~meV), $r_{65}=1.7\, {\rm nm}$ ($E_\textup{VS}=0.75$~meV) in the 1-electron case,
and $r_{5}^{2e}=4.8\, {\rm nm}$ ($E_\textup{VS}=0.58$~meV) in the 2-electron case, consistent with our expectations.

\section*{Supplementary Note 4}

The TCAD model of the real device geometry is shown in Supplementary Figure S2 (a). As mentioned in the main article, in our simulations we use the same voltage range as in the experiments and the only free parameter is the Si/SiO$_2$ interface charge, $Q_\textup{ox}$. This has been chosen to match the experimental threshold voltage ($V_\textup{th}$=0.625 V). In Supplementary Figure S2 (b) the dependence of $V_\textup{th}$ on $Q_\textup{ox}$ is reported. This has been evaluated by integrating the charge density in a region of the 2DEG near the interface as all the gate voltages are simultaneously swept (see inset). The value of choice is $Q_\textup{ox}= -4.5\times〖10〗^{11}$ cm$^{-2}$. The vertical electric field profile in the ($x,z$) plane at $V_\textup{P}$=1.4V is shown in Supplementary Figure S2 (c). As highlighted in the main article, the QD is formed in the region where the conduction band falls below the Fermi energy, so that the vertical interface field therein is the one relevant to calculate the valley splitting. Finally, with the same methodology illustrated in the paper, the dependence of the valley splitting (at $V_P$ = 1.4V) on the interface charge density has been estimated with both the atomistic and the effective mass theories, and is shown in Supplementary Figure S2 (d). It is of note that the variation of $E_\textup{VS}$ with interface charges is relatively small and cannot account for the large offset between computed and experimental values. Hence, we believe that the main factors that reduce valley splitting in this nanostructure are due to interface effects (as discussed in the main article).

\end{document}